\documentclass{osa-article}
\usepackage{siunitx}
\usepackage{textcomp}
\usepackage[utf8x]{inputenc}
\usepackage{multirow}
\journal{oe}

\articletype{Research Article}

\DeclareMathOperator*{\argmin}{argmin}

\begin{document}

\title{Fourier DiffuserScope: Single-shot 3D Fourier light field microscopy with a diffuser}

\author{Fanglin Linda Liu \authormark{1}, Grace Kuo \authormark{1}, Nick Antipa \authormark{1}, Kyrollos Yanny \authormark{2}, and Laura Waller\authormark{1,2,*}}

\address{\authormark{1} Electrical Engineering and Computer Sciences, University of California, Berkeley, 94709, USA\\}

\address{\authormark{2} UCB/UCSF Joint Graduate Program in Bioengineering, University of California, Berkeley, CA, 94709, USA \\}

\email{\authormark{*}waller@berkeley.edu} 



\begin{abstract}

Light field microscopy (LFM) uses a microlens array (MLA) near the sensor plane of a microscope to achieve single-shot 3D imaging of a sample without any moving parts. Unfortunately, the 3D capability of LFM comes with a significant loss of lateral resolution at the focal plane. Placing the MLA near the pupil plane of the microscope, instead of the image plane, can mitigate the artifacts and provide an efficient forward model, at the expense of field-of-view (FOV). Here, we demonstrate improved resolution across a large volume with Fourier DiffuserScope, which uses a diffuser in the pupil plane to encode 3D information, then computationally reconstructs the volume by solving a sparsity-constrained inverse problem. Our diffuser consists of randomly placed microlenses with varying focal lengths; the random positions provide a larger FOV compared to a conventional MLA, and the diverse focal lengths improve the axial depth range. To predict system performance based on diffuser parameters, we for the first time establish a theoretical framework and design guidelines, which are verified by numerical simulations, then build an experimental system that achieves $<\SI{3}{\micro\meter}$ lateral and $\SI{4}{\micro\meter}$ axial resolution over a $1000 \times 1000 \times \SI{280}{\micro\meter \cubed}$ volume. Our diffuser design outperforms the MLA used in LFM, providing more uniform resolution over a larger volume, both laterally and axially.  
\end{abstract}

\section{Introduction}



Volumetric fluorescence microscopy with video-rate capture is essential for understanding dynamic biological systems. Single-shot 3D imaging with a 2D sensor is possible by using a hardware encoding procedure followed by a computational decoding procedure. Light field microscopy (LFM) \cite{lippmann1908LF,okano1997LF,Javidi2004lfm,javidi2006lfm,levoy2006lfm,martinez2018LFMreview} is one popular implementation of this, where a microlens array (MLA) is inserted in front of the microscope's image sensor to simultaneously capture 2D spatial and 2D angular information. The resulting 4D light field can be used for digital refocusing, perspective synthesis, or 3D reconstruction. However, using a 2D sensor to sample a 4D light field requires trading off angular and spatial sampling, resulting in poor resolution. This is particularly undesirable in microscopy, where resolution is the key performance metric.

The resolution of LFM can be improved without requiring multiple measurements~\cite{javidi2003LFsequence} by taking a deconvolution approach to single-shot image
reconstruction~\cite{broxton2013lfm,prevedel2014lfm_worm}. In this case, the captured 2D measurement is used to directly solve for the 3D object, without the intermediate step of calculating a 4D light field. The method makes an implicit assumption of no occlusions, which is valid for most fluorescence microscopy applications. Deconvolution LFM can achieve significantly better (nearly diffraction-limited) resolution at some depth planes, but its performance degrades quickly with depth, even with wavefront coding~\cite{cohen2014enhancing}. Besides suffering from non-uniform resolution, deconvolution LFM incurs artifacts at the native focal plane and requires a computationally-intensive spatially-varying deconvolution procedure. These artifacts and the resolution loss can be mitigated by placing the MLA in an off-focus plane~\cite{lumsdaine2009plenoptic2,li2019plenoptic2LFM,chen2020plenoptic2miniscope}, but the spatial variance and resolution inhomogeneity remain.

To solve some of these problems, an alternative configuration, termed Fourier light field microscopy (FLFM), places the MLA at the Fourier (pupil) plane of the objective, with the sensor one microlens focal length away~\cite{Martnez2016flfm,scrofani2018flfm,guo2019flfm,Martnez2020flfm}. This effectively splits the 2D sensor into a grid of sub-images, with each microlens imaging the sample from a different perspective angle. FLFM achieves more uniform resolution near the native focal plane and has a spatially-invariant point spread function (PSF) for improved computational efficiency. However, the fundamental trade-off between spatial and angular sampling remains (unless a camera array is used, greatly increasing cost and complexity~\cite{lin2015CameraArray}). Previous single-sensor implementations require limiting the microscope's field-of-view (FOV) in order to avoid overlap of the sub-images at the sensor~\cite{Martnez2016flfm,scrofani2018flfm,guo2019flfm,Martnez2020flfm}. The resolution is more homogeneous than LFM, but still degrades quickly with depth.

Our Fourier DiffuserScope improves on FLFM by replacing the regular MLA with a diffuser consisting of randomly-spaced multi-focal microlenses. The new architecture has several advantages: 1) By using microlenses with multiple focal lengths~\cite{Javidi2003multifocusLFM,perwass2012raytrix,georgiev2012multifocusLFM,cong2017rmla}, the PSF will have sharp features at a wide range of depth planes, improving the axial depth range and resolution homogeneity. 2) The randomness of the diffuser eliminates periodicity in the PSF and thus removes the ambiguities that required FOV limits in FLFM. Thus, we allow the microlens sub-images to overlap, then use compressed sensing algorithms \cite{donoho2006compressedsensing,candes2008compressedsensing} to reconstruct the 3D volume without trading off volumetric FOV and depth resolution. This `best of both worlds' scenario is possible only when the sample is sparse in some domain, as is generally true in fluorescence microscopy. The resulting system achieves uniform resolution over a large volume, with imaging speed limited only by signal strength or camera frame rate. 

Fourier DiffuserScope is a variant of our previous methods for diffuser-based imaging with different architectures~\cite{antipa2018diffusercam,antipa2016diffuserLF,liu2019diffuser,monakhova2019learned,kuo2020flatscope, yanny2019randoscope}. Here, we provide the first theoretical framework for Fourier DiffuserScope design with given performance metrics (e.g. resolution, volumetric FOV), and we directly compare with FLFM. We demonstrate the advantages of both the random and multi-focal properties of our diffuser by comparing directly with FLFM and a random uni-focal design. Finally, we build an experimental system, designed in Zemax OpticStudio, that achieves $2$-$\SI{3}{\micro\meter}$ lateral and $\SI{4}{\micro\meter}$ axial resolution over a $1000 \times 1000 \times \SI{280}{\micro\meter \cubed}$ volume. We use the system to record 3D videos of a freely-moving \textit{C. elegans} nematode at 25 fps. 




\section{Related work}


Besides variations of LFM, our work is related to other methods for single-shot 3D fluorescence microscopy.

\textbf{Multifocal microscopy} methods simultaneously capture multiple in-focus images at different depths. This can be done by using beamsplitters and multiple cameras conjugate to different depth planes~\cite{prabhat2004TwoCamera}; however, the resulting system is expensive and bulky. To acquire multiple depths with a single sensor, a distorted phase grating can be inserted in the pupil plane to project different axial layers onto different sub-images on the sensor~\cite{blanchard1999MultifocalGrating,abrahamsson2013multifocus, he2018computational_multifocus}. A similar result can be achieved with superimposed Fresnel lenses~\cite{maurer2010SLMmultifocal} or a diffractive metalens~\cite{guo2019metalens}. For more than a few depth planes, multiplexed volume holography is a good option due to its low cross-talk~\cite{luo2010VolumeHolo}. In all these methods, however, the FOV is sacrificed in order to increase the number of depth planes (generally less than $25$~\cite{he2018computational_multifocus}), limited by how many sub-images fit on the sensor. 

\textbf{PSF engineering for point localization} refers to methods that use a coded mask in Fourier space, like our system, but with the image captured in image (real) space. This results in a depth-dependant PSF (e.g. astigmatic~\cite{kao1994astigmatic,huang2008astigmatic}, double-helix~\cite{greengard2006dhelix1,pavani2009dhelix2,berlich2016dhelix_singleshot}, tetrapod~\cite{shechtman2015tetrapod}, etc.) that, along with inverse algorithms, is well-suited to localize separated point-like molecules \cite{kao1994astigmatic,huang2008astigmatic,greengard2006dhelix1,pavani2009dhelix2,shechtman2015tetrapod}, but ill-posed when the object is continuous~\cite{berlich2016dhelix_singleshot}. Because our Fourier DiffuserScope places both the phase mask and the sensor near the Fourier plane, we have a much larger PSF in which the energy is distributed into more features, so that the cross-correlation of laterally and axially shifted PSFs is lower than that of engineered PSFs. As a result, the design matrix of our random diffuser has nearly orthogonal columns, which is better suited to reconstruct a 3D volume from an undersampled 2D measurement, according to compressed sensing theory~\cite{candes2005rip}.

\textbf{Lensless mask-based imaging}, which uses a coded aperture for lens-free 2D~\cite{asif2015flatcam} or 3D~\cite{adams2017flatscope} imaging, first emerged in X-ray and gamma-ray systems~\cite{dicke1968coded_aper_xray,fenimore1978coded_aper} for 2D imaging in situations where lenses are difficult to implement. Amplitude coded masks are straightforward to design and easy to fabricate, but come with the inherent issue of blocking a lot of light, which leads to low signal-to-noise ratio (SNR) in the acquisition and noise amplification during reconstruction. Phase masks are more difficult to fabricate but have much better light efficiency~\cite{boominathan2020phlatcam}.

\textbf{Diffuser-based microscopy} describes several different architectures that emerged from our original DiffuserCam~\cite{antipa2018diffusercam}, which is a lensless phase-mask-based imager that uses a diffuser for encoding 3D information. We have demonstrated 2D~\cite{kuo2018flat,monakhova2019learned}, 3D~\cite{antipa2018diffusercam,liu2019diffuser,yanny2019randoscope,kuo2020flatscope }, and 4D light field imaging~\cite{antipa2016diffuserLF}, flat~\cite{kuo2020flatscope} and miniature microscopy~\cite{yanny2019randoscope}. Our original diffuser was an off-the-shelf Gaussian pseudo-random phase mask with $100 \%$ fill factor, placed directly in front of the sensor. However, the resulting PSFs had substantial background light, which amplifies noise during deconvolution. Here, we use a designed diffuser made of randomly-located microlenses to focus light into high-contrast random multi-focal spots, providing good SNR across a large depth range, while maintaining the randomness of the PSF.


\section{System overview}
\label{section_overview}
\begin{figure}[htbp!]
\centering\includegraphics[width=\linewidth]{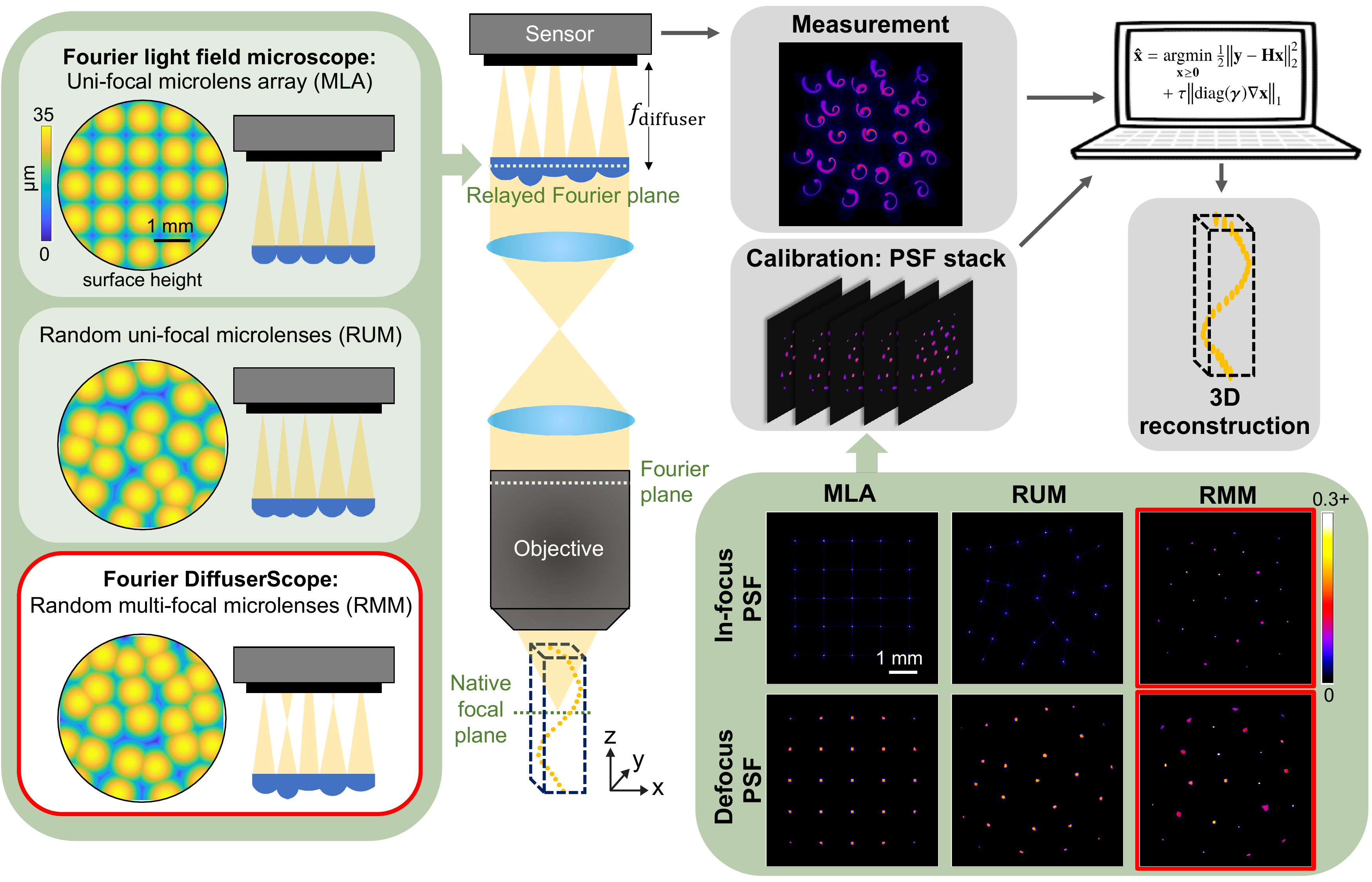}
\caption{System overview for Fourier DiffuserScope and Fourier light field microscopy (FLFM). A diffuser or microlens array is placed in the Fourier plane of the objective (relayed by a 4\textit{f} system) and a sensor is placed one microlens focal length after. From a single 2D sensor measurement, together with a previously calibrated point spread function (PSF) stack, 3D objects can be reconstructed by solving a sparsity-constrained inverse problem. Here, we compare three choices of diffuser/microlens array: FLFM with a uni-focal microlens array (MLA), random uni-focal microlenses (RUM) and our Fourier DiffuserScope with random multi-focal microlenses (RMM). Our RMM design provides a non-periodic PSF with different spots coming into focus at different planes, enabling 3D reconstructions with full FOV and high resolution across a wider depth range. Note that the PSF images (bottom right) are shown with a gamma correction of 0.4 for better visibility.}
\label{fig_system}
\end{figure}

Our Fourier DiffuserScope consists of a diffuser (a phase mask with randomly-located multi-focal microlenses) in the Fourier plane of a microscope objective, with the sensor placed after, spaced by the average focal length of the diffuser (Fig. \ref{fig_system}). Because the actual Fourier plane of the objective is physically inaccessible, we insert a relay system to image its pupil plane onto the diffuser. For each point emitter in the object space, the diffuser will produce a unique multi-spot PSF on the sensor. As compared to Gaussian diffusers or highly-scattered speckle patterns, our diffuser PSF concentrates light onto fewer pixels in order to improve SNR, while also ensuring that different points come into focus at different depths. Because our PSF is distributed and different for each point location within the 3D space, it is possible to reconstruct the whole volume from a single measurement with compressed sensing algorithms. 

To model the image formation process, we divide the 3D volume into 2D slices, where each slice corresponds to a single depth plane. Neighboring slices are separated axially by less than half the axial resolution.  Our experimental system is designed to ensure that the system PSF (the sensor measurement resulting from a single point source) for each depth can be modeled as shift-invariant. Thus, the measurement contribution from each 2D plane is the convolution between the object slice at that depth and the PSF at that depth. The PSFs for different depths have different sizes and different microlenses come into focus, such that each depth has a unique PSF. The final sensor measurement is the sum of the contributions from each 2D layer, assuming that the light from different fluorescent sources is mutually incoherent and there are no occlusions:
\begin{equation}
    \mathbf{y} = \sum_z \mathbf{h}_z * \mathbf{x}_z = \mathbf{Hx}.
    \label{equation_forward}
\end{equation}
Here, $\mathbf{y}$ is the intensity measurement on the sensor, $\mathbf{h}_z$ is the measured PSF at depth $z$ (acquired during calibration), $\mathbf{x}_z$ is the object intensity at depth $z$, and $*$ represents 2D convolution over the transverse dimensions. Since this is a linear operation, we can write our model in matrix form where vector $\mathbf{x}$ is a vector representing the entire 3D volume and $\mathbf{H}$ is a matrix with columns containing the calibration measurements from every depth. Because our system is shift-invariant at each depth, we can compute $\mathbf{Hx}$ computationally efficiently using FFT-based convolutions. The forward model in Eq.~\ref{equation_forward} defines the data fidelity term of our inverse problem. Because we solve for 3D from a single 2D measurement without reducing the number of lateral pixels in the reconstruction, the inverse problem is under-determined. We solve it by using a compressed sensing algorithm that leverages the multiplexed nature of our measurements and assumes the sample is sparse in some domain. This sparsity-constrained inverse solver can be written as:
\begin{equation}
    \mathbf{\hat{x}}= \argmin_{\mathbf{x} \geq \mathbf{0}} \tfrac{1}{2} \big\| \mathbf{y}-\mathbf{Hx} \big\|_2^2 + \tau \big\| \mathrm{diag} (\boldsymbol{\gamma} ) \nabla { \mathbf{x}} \big\|_{1} .
    \label{eq_optimization}
\end{equation}
Here, $\| \cdot \|_2^2$ is the data fidelity term, $\| \cdot \|_{1}$ is a regularization term that enforces sparsity, and $\tau$ is a tuning parameter related to the sparsity level. We use 3D Total Variation (TV) sparsity, with $\nabla= [\nabla_x, \nabla_y, \nabla_z]^\intercal$ being the gradient operator~\cite{rudin1992tv}. Since most fluorophores are isotropic in shape, while the resolution of our system is not necessarily isotropic, we add a weighting vector ${\boldsymbol \gamma}=(\gamma_{xy},  \gamma_{xy}, \gamma_z)$ so that the TV value in the lateral and axial directions are weighted differently. 

The system architecture for our Fourier DiffuserScope is essentially the same as FLFM, except that we use random multi-focal microlenses (RMM) instead of a uni-focal MLA. To demonstrate the advantages of the RMM over MLA, we theoretically and numerically compare performance by looking at the properties of their respective sensing matrices $\mathbf{H}$. To separate out the effects of randomness and multi-focal, we also compare to random uni-focal microlenses (RUM). The MLA (Fig.~\ref{fig_system}) focuses light from the native focal plane into a grid of sharp spots, providing an in-focus PSF with high SNR, but the periodicity causes ambiguity when shifted by more than one pitch, reducing the effective FOV. Randomizing the location of the microlenses (as with the RUM and RMM) breaks the ambiguity, enabling full-FOV imaging with our sparsity-constrained inverse solver. The RUM, like the MLA, uses the same focal length for all microlenses, resulting in blurred PSFs off-focus and therefore a shallow axial imaging range, particularly for high-NA objectives. Assigning different focal lengths to each of the microlenses, as with our RMM (Fig.~\ref{fig_system}), extends the imaging depth range, within which there is always a subset of microlenses in focus. This trades SNR near the native focal plane for an increased depth range due to spreading high-frequency information across the whole volume. The RMM PSFs form nearly orthogonal columns of the design matrix $\mathbf{H}$, enabling a compressed sensing 3D reconstruction with more voxels than there were pixels in the 2D measurement (50$\times$ more in our experimental prototype). 


\section{Diffuser design theory}
\label{section_theory}
\begin{figure}[h]
\centering\includegraphics[width=\linewidth]{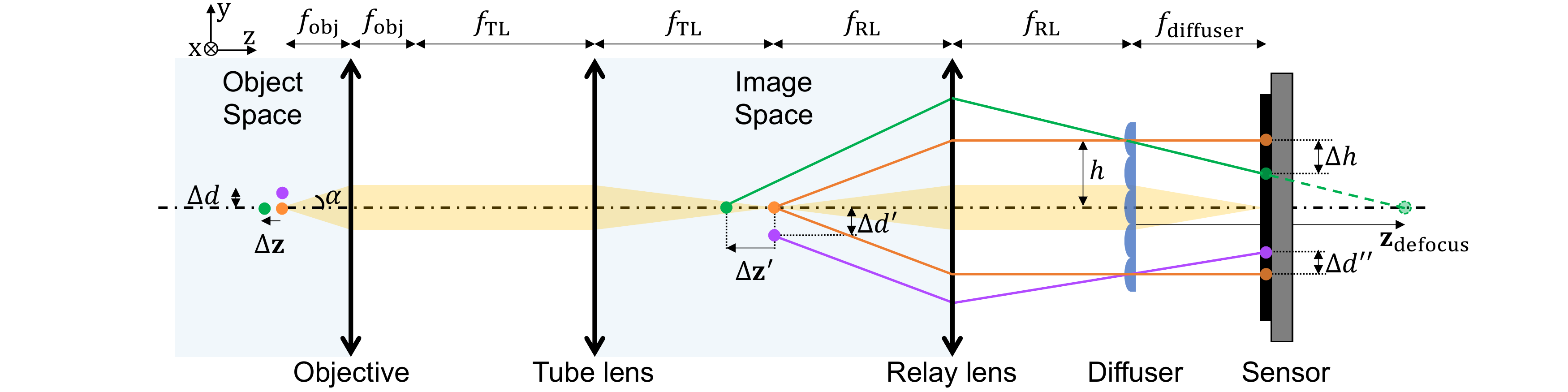}
\caption{System performance analysis. To analyze lateral resolution, consider the orange and purple point sources, laterally spaced by $\Delta d$ in object space, with images on the sensor spaced by $\Delta d''$. For axial resolution, consider the orange and green point sources, axially separated by $\Delta \mathbf{z}$ in object space, which map to their images on the sensor spaced by  $\Delta h$ . The axial resolution is determined by the minimum resolvable separation on the sensor. $\Delta \mathbf{z}$ pointing to the left has a negative value.}
\label{fig_theory}
\end{figure}

In this section, we derive the relationship between diffuser design and system performance in terms of lateral resolution, axial resolution, FOV and depth range. The diffuser is characterized by the following parameters: the average microlens pitch $p$, the number of microlenses on the diffuser $N^2$ (giving an average of $N$ microlenses in each transverse direction), the minimum focal length $f_{\mathrm{min}}$, the maximum focal length $f_{\mathrm{max}}$ and the average focal length $f_{\mathrm{ave}}$ of the microlenses. We compare three different phase masks (MLA, RUM and RMM) to be placed in Fourier configuration. All three designs have the same size and number of microlenses, but the locations and focal-length distributions are different. The MLA and RUM microlenses all have a single focal length $f_{\mathrm{ave}}$, whereas the RMM microlenses all have different focal lengths, varying between $f_{\mathrm{min}}$ and $f_{\mathrm{max}}$. The minimum and maximum focal lengths are designed to focus at the closest and furthest depth planes within the volume-of-interest. The rest focus at depth planes evenly spaced within that range, which means their focal lengths are dioptrically distributed between $f_{\mathrm{min}}$ and $f_{\mathrm{max}}$. The system schematic and parameter definitions are shown in Fig.~\ref{fig_theory} and Table \ref{table_parameter}.

\begin{table}[h]
\begin{center}
\resizebox{\textwidth}{!}{%
\begin{tabular}{|ll|ll|}
\hline

$\lambda$  & optical wavelength   & $N$ & \multirow{2}{*}{\begin{tabular}[t]{@{}l@{}}(average) number of microlenses\\in one dimension\end{tabular}} \\

$n_r$ & refractive index of the medium & & \\

$f_\mathrm{obj}$   & focal length of the objective & $f_\mathrm{max}$ & maximum focal length of RMM   \\

$\mathrm{NA}_\mathrm{obj}$ & numerical aperture of the objective  & $f_\mathrm{min}$ & minimum focal length of RMM   \\

$D$  & diameter of the objective back pupil& $f_{\mathrm{ave}}$ & \multirow{2}{*}{\begin{tabular}[c]{@{}l@{}}(average) focal length of the \\ (multi-focal) diffuser\end{tabular}} \\

$\mathrm{FOV}_\mathrm{obj}$ & diameter of the objective FOV & & \\

$f_\mathrm{TL}$    & focal length of the tube lens & $s$& sensor pixel size \\

$f_\mathrm{RL}$ & focal length of the relay lens & $M$ & system magnification\\
  
$p$ & (average) microlens pitch &  $\mathrm{NA}_\mathrm{eff}$ & effective numerical aperture\\

\hline
\end{tabular}}
\caption{\label{table_parameter} Parameter definitions for the optical system.}
\end{center}
\end{table}

\subsection{Lateral resolution}
\label{section_lres}
In Fourier configuration, each microlens forms a perspective view of the object. Consider the middle microlens in Fig. \ref{fig_theory}, which collects light from the yellow region, with acceptance angle $\alpha$, from an in-focus point source (the orange dot in object space) and forms a diffraction-limited spot on the sensor. Other bundles of light from the same point source will reach other microlenses, focusing to separate spots on the sensor. With the MLA, these spots will form a grid, whereas with the RUM or RMM, they will form a random set of points at the sensor. The in-focus lateral resolution is determined by the size of the diffraction-limited spot beneath a single microlens, which is determined by the effective numerical aperture (NA), or the acceptance angle $\alpha$, of the microlens sub-aperture. Since the the back pupil of the objective is divided into $N$ microlenses in each direction, the effective NA (under paraxial approximation) is the objective NA divided by $N$: $\text{NA}_\text{eff} = \text{NA}_\text{obj} / N $. The diffraction-limited lateral resolution is given by the Rayleigh criterion:

\begin{equation}
\label{eq_lres}
R_\mathrm{lateral} = \frac{1.22 \lambda }{2 \text{NA}_{\text{eff}}}  = \frac{1.22 \lambda N}{2 \text{NA}_{\text{obj}}}  .
\end{equation}
For the RUM and RMM, the aperture size of each microlens varies, so we determine expected resolution by the average sub-aperture size, which is designed to match the MLA effective NA, in order to compare the two situations fairly. 

To achieve the diffraction-limited resolution derived above, the sensor pixel spacing must be small enough to Nyquist sample the pattern after taking into account magnification. To quantify this requirement for our Fourier configuration, consider two point sources laterally separated by $\Delta d$ (the orange and purple dots in the object space of Fig. \ref{fig_theory}). After the 4\textit{f} system of the objective and the tube lens, their intermediate images will be spaced by $\Delta d' = ( f_\mathrm{TL} / f_\mathrm{obj} ) \Delta d $. Using similar triangles between the relay lens, the microlens plane and the sensor, the distance between the two chief rays on the sensor is $\Delta d'' = (f_{\mathrm{ave}} / f_\mathrm{RL} )\Delta d' $. Together, we have $\Delta d'' = M \Delta d $, where $M$ is the lateral magnification rate from the object space to the sensor: 
\begin{equation}
\label{eq_mag}
M =   \frac{f_\mathrm{TL}}{f_\mathrm{obj}} \frac{f_{\mathrm{ave}}}{f_\mathrm{RL}} .
\end{equation}
Thus, the pixel size $s$ satisfies Nyquist sampling when $s \leq M R_\mathrm{lateral} /2$.

Because we reconstruct 3D information, we also investigate how lateral resolution changes for objects away from the objective's native focal plane. For MLA and RUM, in which all microlenses have the same focal length, the minimum resolvable spot is determined by the circle of confusion; we define off-focus lateral resolution to be the radius of the circle of confusion, $\Delta c$. To derive the off-focus resolution in our setup, we first calculate the defocus distance of the intermediate image for an off-focus point source (the green dot in object space in Fig. \ref{fig_theory}), which is scaled by the objective's magnification: $\Delta \mathbf{z}' = ( f_\mathrm{TL} / f_\mathrm{obj} )^2 \Delta \mathbf{z}$. Then, by applying the Newtonian form of the thin lens equation for the relay lens, we calculate the location of the second intermediate image of the green point, relative to the diffuser, after passing through the relay lens:  $ \mathbf{z}_\mathrm{defocus} = -f_\mathrm{RL}^2 /\Delta \mathbf{z}'$. This serves as the `object' for the diffuser microlenses and $\mathbf{z}_\mathrm{defocus}$ is the `object distance'. So, the circle of confusion size depends on $\mathbf{z}_\mathrm{defocus}$, the diffuser focal length $f_{\mathrm{ave}}$ and the size of a single microlens $p$. The resulting expression describes how the lateral resolution degrades linearly with defocus distance:
\begin{equation}
\label{eq_circle_confusion}
\Delta c=  \frac{p \, f_{\mathrm{ave}}}{2}   \frac{1}{| \mathbf{z}_\mathrm{defocus} |} = \frac{p \, f_{\mathrm{ave}}}{2}   \frac{f_\mathrm{TL} ^2 |\Delta \mathbf{z} |} { f_\mathrm{RL}^2 f_\mathrm{obj}^2} .
\end{equation}

The primary advantage of the RMM that we use in Fourier DiffuserScope is that it has multiple focal lengths. Thus, the subset of microlenses that are in focus at each depth will have spots with size matching the in-focus diffraction-limited lateral resolution derived in the previous section. Hence, the lateral resolution does not degrade with depth within the volume-of-interest. When the object moves beyond the designed range, the lateral resolution will degrade linearly with defocus distance.  A detailed analysis on the depth range is in Sec. \ref{section_depthrange}.

\subsection{Axial resolution}
\label{section_ares}
We define the axial resolution as the minimum axial distance between two point emitters that can be resolved in the reconstruction. The off-axis microlenses have disparity, such that point sources from different depths are imaged to different lateral locations on the sensor; two points will be resolved if their images are separated by at least the diffraction-limited lateral resolution (after magnification). We analyze the limits for the outermost microlens, which has the largest disparity angle. In Fig. \ref{fig_theory}, the center of the topmost microlens is $h = (N-1) /2 \cdot p$ away from the optical axis. Two point sources with the same lateral location are axially spaced by $\Delta \mathbf{z}$ (the orange and green dots in object space, Fig. \ref{fig_theory}). In the previous section we have already related $\Delta \mathbf{z}$ to $\mathbf{z}_\mathrm{defocus}$. From the similar triangles formed by the relay lens, the diffuser and the sensor, we can calculate the lateral distance between the orange chief ray and the green chief ray on the sensor, $\Delta h = (f_{\mathrm{ave}} / | \mathbf{z}_\mathrm{defocus} |) h $. The minimum distance on the sensor for resolving the points is $MR_\mathrm{lateral}$, which sets the minimum value for $\Delta h$, and the value of  $\Delta \mathbf{z}$ we solve for is the axial resolution $R_\mathrm{axial}$. Given the relation between the relayed pupil diameter and numerical aperture $N \cdot p =( f_\mathrm{RL} / f_\mathrm{TL})2 \mathrm{NA}_\mathrm{obj} f_\mathrm{obj}$, the axial resolution is:
\begin{equation}
\label{eq_ares}
R_\mathrm{axial} =\frac{N}{N-1} \frac{1}{ \mathrm{NA}_\mathrm{obj} } R_\mathrm{lateral} =  \frac{N^2}{N-1} \frac{1.22  \lambda}{2 \mathrm{NA}_\mathrm{obj}^2}  .
\end{equation}
The axial resolution off-focus is determined in a similar way, except that the two point images must now be separated by a distance of at least the circle of confusion size. We derive this by replacing the $R_\mathrm{lateral}$ term in Eq. \ref{eq_ares} with the radius of the circle of confusion $\Delta c$ in Eq. \ref{eq_circle_confusion}. Note that the slope of the defocused axial resolution as a function of depth is proportional to that of defocused lateral resolution.

\subsection{Field-of-view}
\label{section_fov}
We analyze the in-focus FOV for each of the three microlens designs, and assume that the FOV throughout the volume will be approximately the same as that at the native focal plane. The regular layout of the MLA results in a periodic PSF. When a point in the scene moves laterally by an amount that shifts the PSF by an integer number of pitches, the shifted PSF is nearly the same as the unshifted one; this creates ambiguities that cause the deconvolution to fail. To avoid this problem, a field stop is inserted to guarantee that the PSF shifts by less than one period over the FOV~\cite{scrofani2018flfm,guo2019flfm}. The resulting FOV for the MLA-based FLFM is thus limited by the microlens pitch size: $ \mathrm{FOV}_\mathrm{MLA} = p / M$. 

The randomly located lenses in the RUM and RMM create PSFs with randomly-located spots that do not suffer from the ambiguity caused by periodicity. So, both RUM and RMM are able to reconstruct images with the full objective FOV, giving $ \mathrm{FOV}_\mathrm{RUM} = \mathrm{FOV}_\mathrm{RMM} = \mathrm{FOV}_\mathrm{obj}$. This is based on ideal optics; in reality, aberrations can break the shift-invariance of the PSF in the peripheral FOV so that the final reconstruction has a smaller FOV or reduced resolution near the edges. In practice, we determine the FOV for by calculating the similarity between on-axis and off-axis PSFs, described in more detail in Sec. \ref{section_simulation_fov}. 

\subsection{Depth range}
\label{section_depthrange}
The depth range describes the axial distance over which the object can be reconstructed with diffraction-limited resolution. For the uni-focal designs (MLA and RUM), the depth range is simply the depth-of-field (DOF) of a single microlens, since all microlenses have the same focal length. The microscope DOF expression is the sum of a wave optics term and a geometrical optics term~\cite{microscopyU_DOF}, and we use the effective NA to account for the partitioning of the back pupil plane into multiple microlenses:
\begin{eqnarray}
\label{eq_dof}
\mathrm{DOF_{microlens}} &=& \frac{\lambda n_r}{\text{NA}_{\text{eff}}^2} + \frac{n_r \cdot s}{M \cdot \text{NA}_{\text{eff}}}  .
\end{eqnarray}
The main advantage of using multi-focal microlenses in the RMM for Fourier DiffuserScope is that the depth range will be much larger, since the DOFs of different microlenses are offset. The RMM can be designed for the largest possible depth range by ensuring that the focus positions of different microlenses are separated axially by their DOF; thus, the upper bound of the depth range is the product of a single microlens' DOF and the number of microlenses, $N^2 \times \mathrm{ DOF_{microlens}}$.

To design such a RMM to cover a depth range from $- z$ to $+ z$, the maximum and minimum focal lengths are designed to focus on the farthest and nearest depth planes: $1/f_{\mathrm{max}} = 1/f_{\mathrm{ave}} - (f_\mathrm{TL}/(f_\mathrm{RL} f_\mathrm{obj}))^2 (- z)$ and $1/f_{\mathrm{min}} = 1/f_{\mathrm{ave}} - (f_\mathrm{TL}/(f_\mathrm{RL} f_\mathrm{obj}))^2 z$. The remaining focal lengths are dioptrically distributed between $f_\mathrm{min}$ and $f_\mathrm{max}$ to provide equally-spaced focus planes in the object space. In practice, because the microlenses have different sizes and shapes, there will be variation in the resolution of different microlenses. To ensure stability of performance, we design the DOFs to overlap by half, such that the depth range covers half of its upper bound.


\section{Simulation results}
\label{section_simulation}
We use simulations to numerically validate the design theory derived in the previous section and to demonstrate the advantage of using RMM over MLA and RUM. We set the target performance to be $\sim \SI{2}{\micro\meter} $ resolution across a $\sim \SI{200}{\micro\meter} $ depth range using a $20\times$, 1.0NA objective lens ($f_\mathrm{obj}= \SI{9}{\milli\meter}$, $ \mathrm{NA_{obj}}=1.0$, $\mathrm{FOV_{obj}}=\SI{1.1}{\milli\meter}$, $D=\SI{18}{\milli\meter} $). The design wavelength is $\lambda = 510 \ \mathrm{nm}$ for common green fluorescent calcium indicators. The tube lens and the relay lens form a $1:1$ relay system to conjugate the back pupil plane onto the phase mask, so the diffuser side length equals the pupil diameter ($ N \cdot p =  \SI{18}{\milli\meter} $). Calculated from Eq. \ref{eq_lres} and Eq. \ref{eq_ares}, the diffuser has at most $N=5$ microlenses in one transverse direction with an average pitch size $p=\SI{3.6}{\milli\meter}$, resulting in an effective NA of $0.2$ and predicted resolution $R_\mathrm{lateral} = \SI{1.56}{\micro\meter}$ and $R_\mathrm{axial}=\SI{1.94}{\micro\meter}$. The average focal length of the RMM ($f_{\mathrm{ave}}= 58.5 $ mm), matched to the focal lengths of the MLA and RUM, is chosen to achieve a total magnification of $M=6.5 \times$.  For the RMM, with our goal of $\pm z = \SI{\pm 100}{\micro\meter}$, the microlens focusing at the nearest and farthest depth planes have $f_\mathrm{min}= \SI{54.6}{\milli\meter}$  and $f_\mathrm{max}= \SI{63.1}{\milli\meter}$, respectively. The focal lengths of the remaining $23$ microlenses are dioptrically distributed between $f_{\mathrm{min}}$ and $f_{\mathrm{max}}$. 

The surface height of the three phase mask designs are shown in Fig. \ref{fig_system}. The centers of the randomly-spaced microlenses are generated from a uniform distribution, under the constraint that the distance between adjacent centers is at least $70 \%$ of the microlens pitch. A spherical surface is placed at the center of each microlens location, taking into account the focal lengths and refractive index ($n_r = 1.56$ for photopolymer). Then, we take the point-wise maximum surface height to form the final diffuser with $100 \%$ fill factor. The sensor is located at the distance of the average focal length behind the diffuser, with our $\SI{2}{\micro\meter}$ pixel size ensuring Nyquist sampling of the diffraction-limited pattern. 

Our simulation framework models wave-optical propagation from the object to the sensor. From a point source location in the object space, we calculate the spherical wavefront at the the back focal plane, then multiply by the apodization function of the objective to get the wavefront distribution at the pupil plane. The wavefront at the pupil is then scaled by the relay system, multiplied by the transmission function of the diffuser/MLA, then propagated to the sensor using the angular spectrum method~\cite{goodman2005FourierOptics}. The resulting PSFs are shown in Fig. \ref{fig_system}, with the in-focus PSF being the intensity pattern at the sensor for a point source at the native focal plane of the objective, and the defocus PSFs being the intensity for a point source off-focus by $\SI{100}{\micro\meter}$ towards the objective. For both uni-focal designs (MLA, RUM), all the lenslets are in-focus or out-of-focus simultaneously, while for RMM each microlens comes into focus at a different plane. 

\subsection{Resolution}
\begin{figure}[h!]
\centering\includegraphics[width=\linewidth]{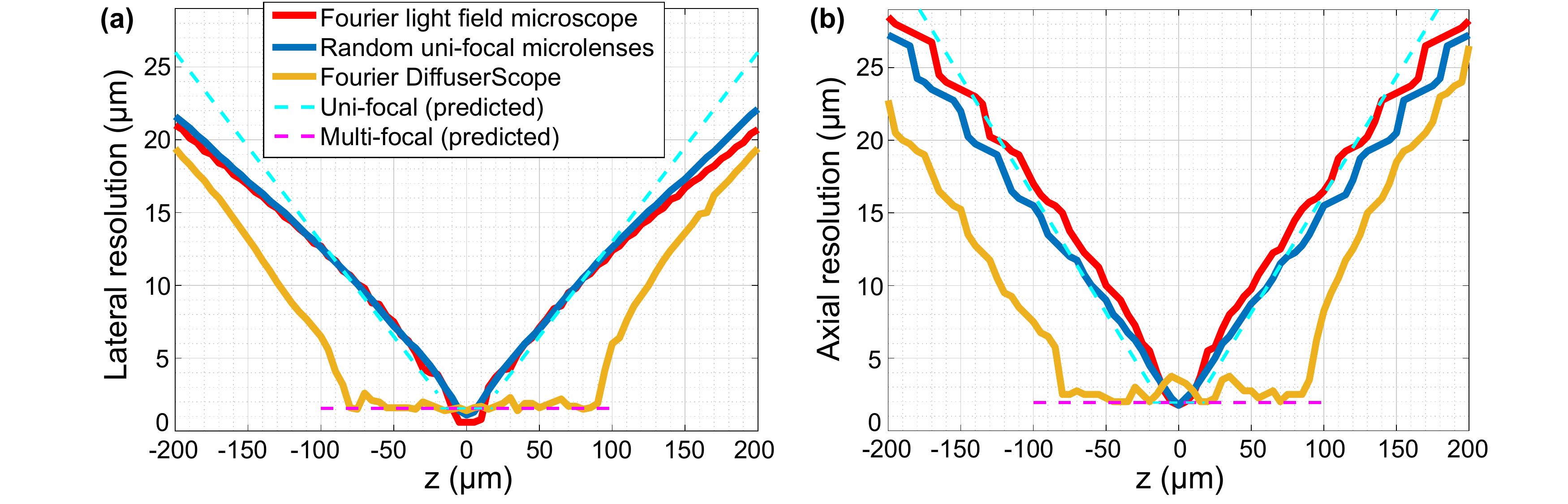}
\caption{Comparison of simulated and theoretical two-point resolution at different depth planes for the three cases of Fourier microlens designs: Fourier light field microscope's MLA, RUM and our Fourier DiffuserScope's RMM. (a) Lateral resolution and (b) axial resolution at different depth planes. The MLA used in Fourier light field microscope (red solid line)  and the random uni-focal microlenses (blue solid line) have the best resolution at the native focal plane ($z=0$) but the performance degrades rapidly outside a small range of depth planes ($z=\SI{-10}{\micro\meter}$ to $z=\SI{10}{\micro\meter}$), as predicted by theory (cyan dashed line). The RMM used in our Fourier DiffuserScope (orange solid line) has slightly worse resolution at $z=0$, but achieves good resolution across a much larger depth range ($z=\SI{-80}{\micro\meter}$ to $z=\SI{90}{\micro\meter}$). Within this range, the resolution stays fairly close to the predicted multi-focal resolution (magenta dashed line). }
\label{fig_res}
\end{figure}

To characterize the lateral and axial resolution, which vary with depth, we reconstruct volumes from acquisitions with two point sources at varying separation distances. For this very sparse scene, the reconstruction converges in only 5 iterations of Richardson-Lucy without regularization~\cite{richardson1972rl,lucy1974rl}. We stop after 8 iterations and consider the two points resolved when there is at least a $20 \% $ intensity drop between them, as in the Rayleigh criterion. For lateral resolution, the two points are placed on the same depth plane with separation only in the $x$-$y$ direction; for axial resolution, the two points are both on the optical axis and symmetrically set apart from the designated depth plane. The results in Fig. \ref{fig_res} compare the reconstruction resolution for the three diffuser/MLA designs, with comparison to the theory presented in Sec \ref{section_theory}.

\label{section_simulation_res}

At the native focal plane ($z=\SI{0}{\micro\meter}$) the MLA has a lateral resolution of $\SI{0.6}{\micro\meter}$ and the RUM has a lateral resolution of $\SI{1.1}{\micro\meter}$ (Fig. \ref{fig_res}(a)), somewhat better than the predicted $R_\mathrm{lateral}= \SI{1.56}{\micro\meter}$ owing to deconvolution. However, the resolution of both uni-focal designs (MLA, RUM) degrades rapidly with depth; based on Eq. \ref{eq_circle_confusion}, the slope of the resolution with depth is 0.13 laterally and 0.1625 axially. The lateral resolution of our Fourier DiffuserScope (RMM) remains relatively steady over a large depth range ($z=\SI{-80}{\micro\meter}$ to $z=\SI{90}{\micro\meter}$), varying between $1.4 \sim \SI{2.6}{\micro\meter}$. 

The axial resolution (Fig. \ref{fig_res}(b)) follows similar trends. The highest axial resolution for both MLA and RUM is $\SI{1.75}{\micro\meter}$ at the native focal plane, which is somewhat better than our theoretical prediction of $R_\mathrm{axial}=\SI{1.94}{\micro\meter}$ (Sec. \ref{section_ares}). The axial resolution of RMM oscillates between $2.0 \sim \SI{3.8}{\micro\meter}$ within a $\SI{170}{\micro\meter}$ depth range. Thus, we conclude that our RMM design, relative to the MLA and RUM, slightly sacrifices lateral and axial resolving power at the native focal plane, but gains uniformly high performance across a large imaging volume.

\subsection{Field-of-view}
To compare the FOV of the three different designs, we simulate and reconstruct a 2D phantom that fills the objective FOV ($1.1 \times \SI{1.1}{\milli\meter \squared}$), placed at the native focal plane of the objective (where the uni-focal microlenses have the best performance). The theory in Sec.~\ref{section_fov} predicts that the random diffusers (RUM, RMM) should be able to reconstruct the whole object, while the MLA will only reconstruct $\mathrm{FOV}_\mathrm{MLA} = \SI{554}{\micro\meter}$.

\label{section_simulation_fov}
\begin{figure}[h!]
\centering\includegraphics[width=\linewidth]{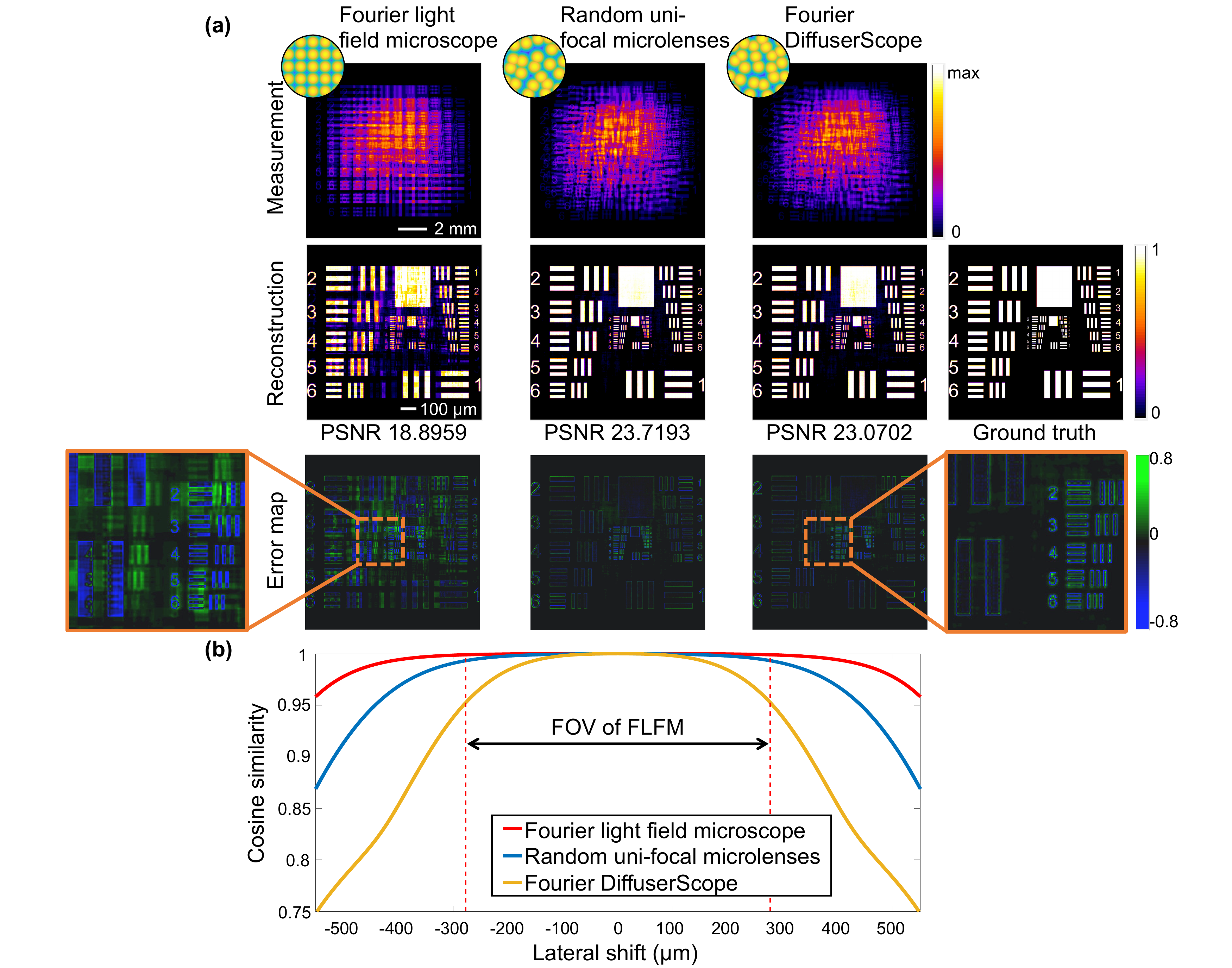}
\caption{Simulations comparing field-of-view (FOV) for different microlens designs. (a) The FLFM (with MLA) reconstruction suffers from ghosting replicas (green regions in the error map) due to its periodic structure. Both the RUM and the RMM reconstruct the phantom successfully. The error of the random diffusers mainly occurs at sharp edges, which can be fixed by adding total variation regularization. Error = reconstruction $-$ ground truth. (b) Cosine similarity between the on-axis PSF and off-axis PSFs is used to quantify the shift-invariance assumption. The MLA has the highest similarity value (red), but its FOV is limited by the microlens pitch. The similarity of RUM (blue) and RMM (orange) are all above $75 \%$ across the full objective FOV. }
\label{fig_fov}
\end{figure}

To simulate the imaging pipeline accurately, we take into account the aberration from plano-convex microlenses, which means that the PSF at the edges of the FOV will have subtle differences from the center PSF. We divide the object into $10 \times \SI{10}{\micro\meter \squared} $ blocks, convolve each block's content with its corresponding PSF (calculated at the center of the block) and then sum up the convolution result from all the blocks to get the simulated measured image. After the spatially-variant block-wise convolution is done, we add $5 \%$ Gaussian noise to generate the final measurement shown in Fig. \ref{fig_fov}(a), first row. The simulated MLA measurement has a periodic pattern because of its periodic PSF, while the diffuser measurements are more random.

To reconstruct the image, we deconvolve the simulated acquisition with a single on-axis PSF (assuming shift invariance) using Richardson-Lucy deconvolution~\cite{richardson1972rl,lucy1974rl}. The result is shown in Fig. \ref{fig_fov}(a), second row. No regularization is added ($\tau=0$ in Eq. \ref{eq_optimization}) in order to compare the worst-case performance. The reconstruction using the MLA shows periodic replicas and large errors, due to the ambiguity of its PSFs. Restricting the FOV with a field stop eliminates this ambiguity at the cost of a reduced FOV. Both random diffusers, which do not have ambiguities in their PSFs, are able to reconstruct the whole object faithfully. The RUM has a slightly better peak signal-to-noise ratio (PSNR), since at the native focal plane all its microlenses are in focus, while only some of the RMM microlenses are. The error maps (error = reconstruction $-$ ground truth) in Fig. \ref{fig_fov}(a) show significantly less error for the random microlenses designs than for the MLA, and errors for the random designs are mainly at edges of objects, which can be reduced by adding TV regularization to the reconstruction.

The shift-variance introduced by the aberrations in our simulation will cause model-mismatch that reduces the performance of the system when using a single-PSF reconstruction. To quantify the shift-variance, we examine the cosine similarity (normalized cross-correlation) between the on-axis and off-axis PSFs (Fig. \ref{fig_fov}(b)). At each lateral shift location, we register the off-axis PSF to the on-axis PSF \cite{guizar2008registration} and calculate the inner product between them. The similarity value for randomly-located microlenses is at least $75 \%$ across the FOV, which is sufficient for single-PSF deconvolution~\cite{antipa2018diffusercam}. At the edges of the FOV, the similarity goes down because the aberration and distortion are most severe at the periphery. The MLA provides the highest values because all microlenses have a regular shape and are of the same size, but the benefits are not useful because the FOV is actually limited by periodicity, as described in Sec. \ref{section_fov}. The randomly distributed microlenses have irregular borders where the surfaces of neighboring microlenses are merged, which increases the aberration, and the multi-focal diffuser adds additional defocus aberration as compared to the uni-focal diffuser.

If high accuracy near the periphery is important, we can correct model mismatch with a spatially-varying deconvolution algorithm~\cite{kuo2020flatscope}. This algorithm calibrates the PSFs at multiple points across the FOV and interpolates them to find the PSFs at each position. It should give better reconstructions, but at a cost of significantly longer computation times and larger memory requirements. In our experimental system, the highest angle incident onto the diffuser (13 degrees) is much smaller than the highest angle (50 degrees) in \cite{kuo2020flatscope}, and the shift-invariant assumption holds well. Thus, we choose to use only a single PSF for each depth for computational efficiency.

\subsection{Depth range}

The two-point resolution in Fig. \ref{fig_res} can be used to estimate the depth range. For the uni-focal designs (MLA, RUM), the lateral resolution remains below its predicted in-focus value over a range of $\sim \SI{20}{\micro\meter}$, which is in agreement with the depth range predicted by Eq. \ref{eq_dof} using our system parameters: $\mathrm{DOF}_\mathrm{microlens} = \frac{0.51 \times 1.33}{0.2^2} + \frac{1.33 \times 2}{6.5 \times 0.2} = \SI{19}{\micro\meter}$. The multi-focal design has stable performance from $z=\SI{-80}{\micro\meter}$ to $z=\SI{90}{\micro\meter}$, demonstrating the improvement of depth range over uni-focal designs.

\label{section_simulation_dr}
\begin{figure}[h!]
\centering\includegraphics[width=\linewidth]{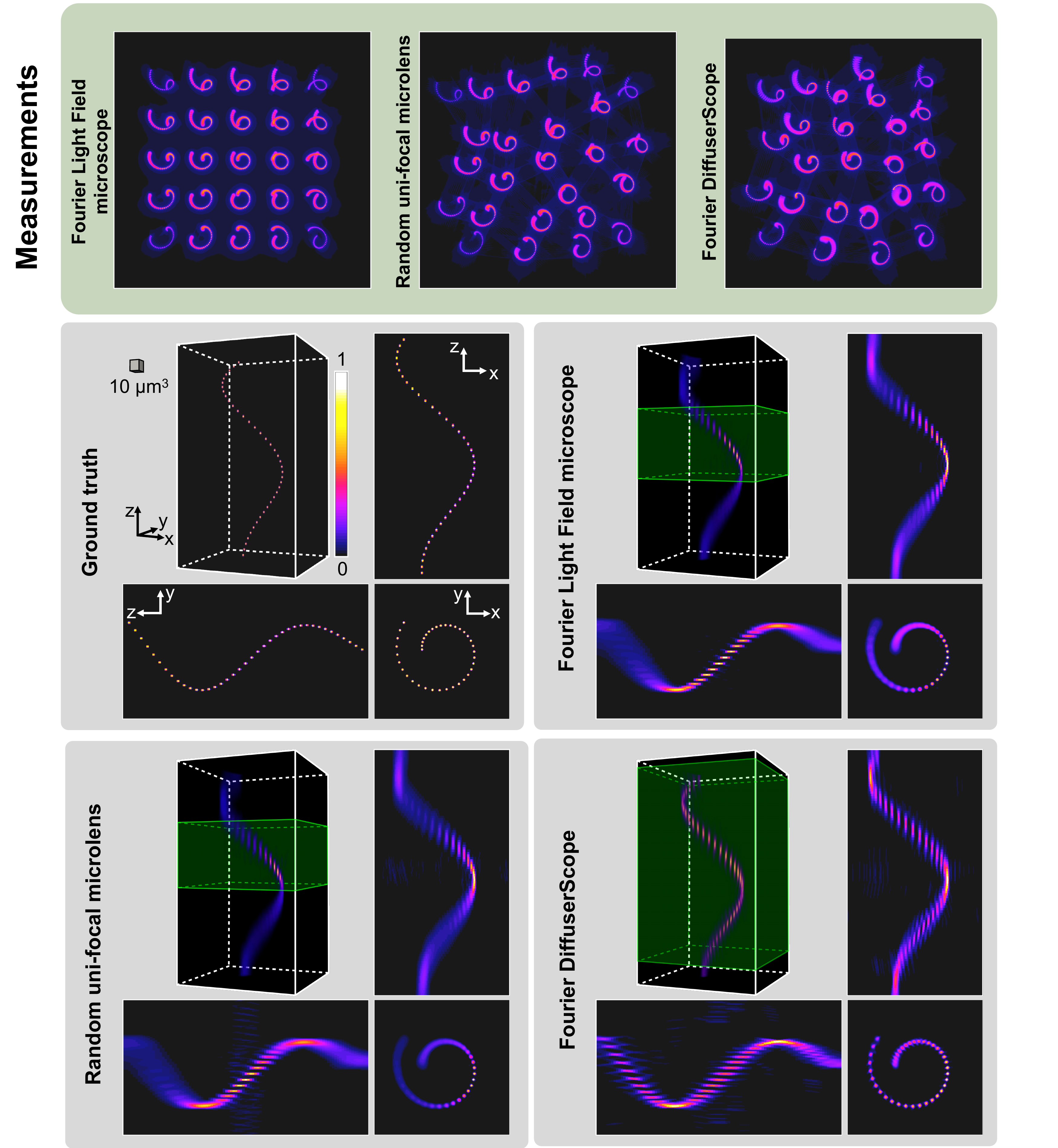}
\caption{Simulated 2D measurements and 3D reconstructions of a sparse spiral object with different microlens designs. The ground truth object is a $\SI{200}{\micro\meter}$-long spiral made of spheres. The Fourier light field microscope (MLA) and the RUM only resolve the spheres in the area around the native focal plane (green shaded area), whereas our Fourier DiffuserScope (RMM) extends the depth range to cover almost the entire volume. }
\label{fig_depth}
\end{figure}

To demonstrate the depth range differences, we reconstruct a long 3D spiral of point sources covering a $\SI{200}{\micro\meter}$ depth range (Fig. \ref{fig_depth}). This phantom contains 39 spheres of $\SI{2}{\micro\meter}$ diameter, with the first one at $z=\SI{95}{\micro\meter}$ and the last one at $z=\SI{-95}{\micro\meter}$, spaced axially by $\SI{5}{\micro\meter}$ (resulting in a  $\SI{3}{\micro\meter}$ gap between spheres axially). The lateral distance between the spheres starts from $\SI{3}{\micro\meter}$ (gap is $\SI{1}{\micro\meter}$) at the center of the spiral and increases up to $\SI{7}{\micro\meter}$ (gap is $\SI{5}{\micro\meter}$) at the outer circle of the spiral. The lateral extent of the spiral ($\SI{66}{\micro\meter}$) stays within the restricted FOV of the MLA to avoid ghosting artifacts. We divide the $\SI{200}{\micro\meter}$-long object into 200 layers of 2D slices, implement the forward model in Eq. \ref{equation_forward} and add $5 \%$ Gaussian noise to the simulated measurement (Fig. \ref{fig_depth}). The measurement contains $25$ sub-images of the spiral object, one for each microlens which observes the spiral from a specific angle; in this way the 3D information is encoded into a single 2D acquisition. The simulated measurements highlight why the depth range of the multi-focal RMM (Fourier DiffuserScope) is much larger than the uni-focal design cases. For the uni-focal (MLA, RUM) cases, only the waist area of the spiral is sharp in all the sub-images. For the multi-focal RMM, different spiral sub-images contain different sharp areas; hence, more in-focus information about the entire depth range passes into the measurement. 

The 3D reconstructions for each of the three cases are shown in Fig. \ref{fig_depth}. We use a PSF calibration stack with fewer (100) PSFs than were used in the forward simulation, to mimic practical axial sampling rates of continuous objects. The sparsity parameter ($\tau =10^{-5}$) is hand-tuned and remains the same for all cases. From the reconstructions, the benefit of using multi-focal microlenses is obvious. 36 spheres are clearly resolved (from $z=\SI{-80}{\micro\meter}$ to $z=\SI{95}{\micro\meter}$) with the RMM design, while only up to 13 spheres are resolved with the uni-focal designs (green shaded regions). The depth range of the three cases matches the depth range where the axial two-point resolution is under $\SI{5}{\micro\meter}$. However, from both the two-point resolution result and the 3D object reconstruction result, the depth range of our RMM is slightly worse than predicted. This is likely due to most microlenses being out-of-focus at both ends of the targeted depth range, causing a lack of high-frequency information that is difficult to deconvolve.

\section{Experimental system and results}
We build an experimental Fourier DiffuserScope system using the RMM design from our simulation, with a $20\times$, 1.0 NA objective lens (Olympus XLUMPLFLN). The fluorescent sample is excited with blue light from a Xenon lamp light filtered by a band-pass emission filter (Semrock FF01-474/27). The emitted green light is filtered using a dichroic mirror (Semrock FF495-Di03)  and an emission filter (Semrock FF01-520/35). Since the back pupil diameter is larger than the sensor size (Andor Zyla 4.2, sensor size 13.3 $\times$ 13.3 mm, pixel size $\SI{6.5}{\micro\meter}$), we demagnify the pupil by 3.75$\times$ so that the full FOV can be recorded. The relay lens design is optimized using Zemax OpticStudio to reduce aberration (see Supplemental materials). 

\label{section_experiment}

\begin{figure}[h!]
\centering\includegraphics[width=\linewidth]{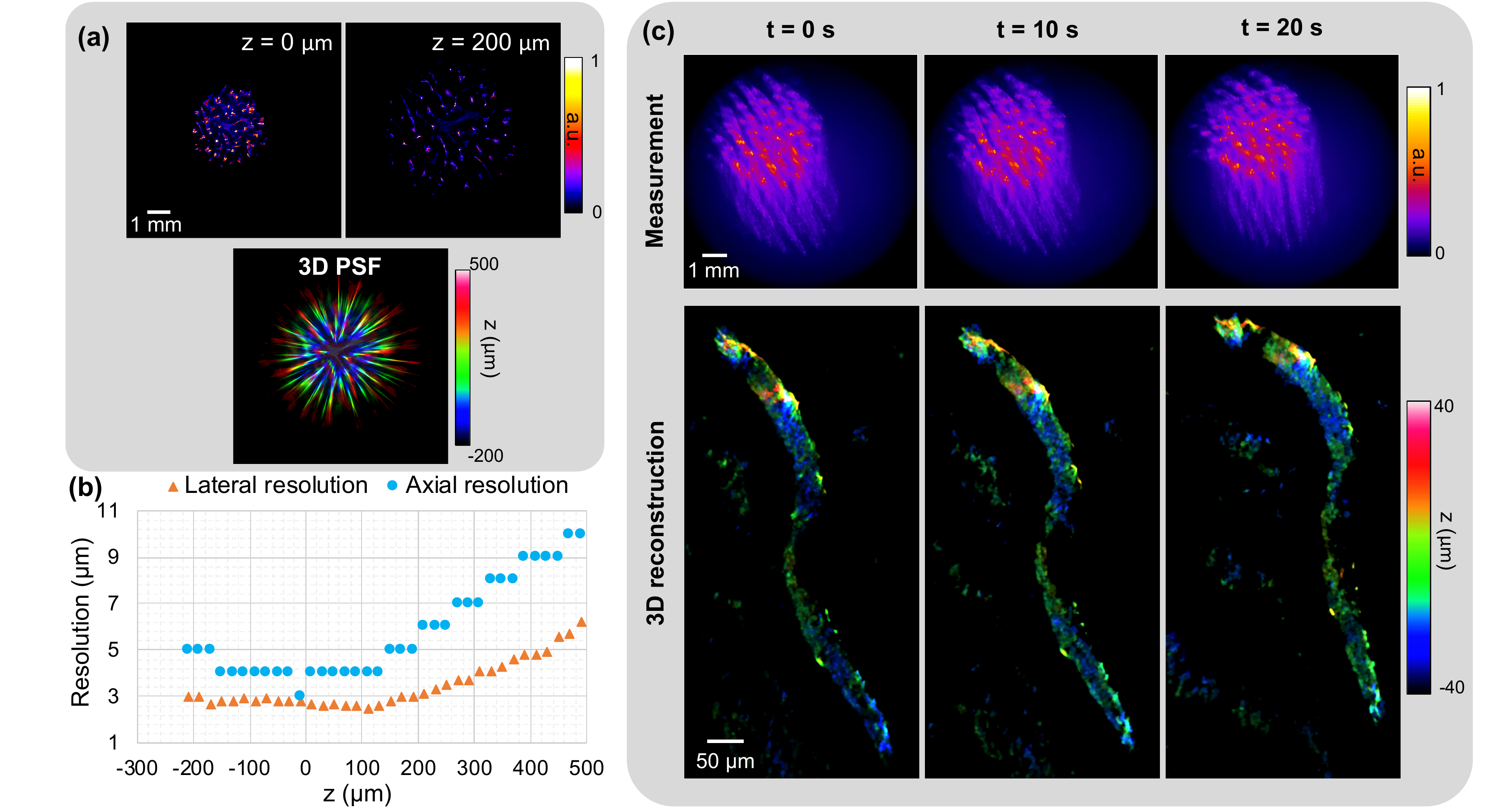}
\caption{Experimental results. (a) Two sample PSFs measured with a point source at $z= \SI{0}{\micro\meter}$ and $z= \SI{200}{\micro\meter}$ depth planes, as well as the 3D PSF plotted with different depth planes color-coded according to the color bar. (b) Experimentally measured resolution, defined as the minimum separation distance at which two sub-resolution fluorescent beads are resolvable. Across the $\SI{280}{\micro\meter}$ depth range from $z=\SI{-150}{\micro\meter}$ to $z=\SI{130}{\micro\meter}$, the lateral resolution is $< \SI{3}{\micro\meter}$ and the axial resolution is less than or equal to $\SI{4}{\micro\meter}$. (c) Raw measurements and 3D reconstructions of a GFP-tagged \textit{C. elegans} recorded at 25 fps. The depth across a \SI{80}{\micro\meter} range is color-coded according to the color bar. The full video is in Visualization 1.}
\label{fig_worm}
\end{figure}

To fabricate our RMM diffuser, we make a negative mold by randomly indenting polished copper using ball bearings with diameters ranging from 10 mm to 16 mm. We then use polydimethylsiloxane (PDMS) to make a replica of the mold with convex-plano microlenses. This approximately achieves our diffuser design parameters of average focal length of 15.6 mm (after considering the relay system), with minimum and maximum focal lengths of 12.3 mm and 21.4 mm, respectively, giving a $\sim \SI{200}{\micro\meter}$ depth range. The main fabrication errors come from deformation error during indentation and shrinkage of the PDMS material. These have opposite effects, since the indented deformation will have bigger diameter than the indenter while the material shrinkage gives smaller diameter, so they offset each other to some extent. 

Fabrication errors should be accounted for during calibration, such that they do not cause model mismatch during deconvolution. The calibration PSF stack is acquired by placing a sub-resolution fluorescent bead at different depths, controlled by a motorized stage, then recording the intensity with a sensor located at the average back focal plane of the diffuser. In total, 350 PSF images were recorded with a $\SI{2}{\micro\meter}$ axial increment from $\SI{200}{\micro\meter}$ below the native focal plane to $\SI{500}{\micro\meter}$ above. When the point source moves more than $\SI{200}{\micro\meter}$ below the native focal plane, the overall PSF becomes so small that the out-of-focus blur from neighboring microlenses will merge into each other, which causes very noisy reconstruction, so we avoid placing objects in this region. Based on the PSF measurements, shown in Fig. \ref{fig_worm}a, there are $\sim 60$ microlenses in the illuminated region of the diffuser. We apply the theory in Sec. \ref{section_lres} and \ref{section_ares} to get the predicted lateral resolution of $\SI{2.4}{\micro\meter}$ and axial resolution of $\SI{2.8}{\micro\meter}$.

We experimentally measured the two-point resolution  (Fig. \ref{fig_worm}b) in order to benchmark the resolution and depth range of our Fourier DiffuserScope prototype. For practicality, measurements were synthetically generated by summing two experimental PSFs at different locations. The image was recovered by solving Eq. \ref{eq_optimization}, and we then calculated the smallest distance at which the two points were still resolved, both laterally and axially, at each depth $z$. The increment of separation distance is $\SI{0.1}{\micro\meter}$ laterally and $\SI{1}{\micro\meter}$ axially. Across a depth range of $\SI{280}{\micro\meter}$  (from $z= \SI{-150}{\micro\meter}$ μm to $z = \SI{130}{\micro\meter}$), the lateral resolution fluctuates between $\SI{2.5}{\micro\meter}$ and $\SI{2.9}{\micro\meter}$ and the axial resolution is mostly $\SI{4}{\micro\meter}$, close to their theoretical predictions. The depth range is larger than the design, suggesting that the actual diameter range of the microlenses is wider than the ball bearings used.

We next demonstrate our system on a live adult \textit{C. elegans} organism that is pan-neuronally expressing a GCaMP fluorescent indicator. The \textit{C. elegans} is anesthetized by levamisole in M9 buffer and then loaded into a $1000 \times 1000 \times \SI{100}{\micro\meter \cubed}$ arena on a microfluidic chip which constrains the worm to move within the FOV of the objective. Since our method is able to reconstruct a 3D object from a single shot, the frame rate is only limited by the sensor. We recorded a raw video at 25 fps while the worm was freely moving (Fig. \ref{fig_worm}c). There is one \textit{C. elegans} image behind every microlens and in total we see $\sim 60$ overlapping \textit{C. elegans} images, each from a different angle. Given that every location in the object space has a unique PSF on the sensor, we are able to deconvolve the overlapping images. Our deconvolution algorithm applies ADMM due to its fast convergence rate \cite{boyd2011ADMM}. To save memory, we did not deconvolve the measurement with all the calibrated PSFs, instead we firstly use a coarse axial sampling of the PSFs to locate the object occupied depth range and then a small subset of fine sampling PSFs to reconstruct the object. The \textit{C. elegans} in our raw video moves within a $\SI{80}{\micro\meter}$ depth range and we use 50 PSFs with $\SI{2}{\micro\meter} $ axial increment to cover the whole object. The reconstructed \textit{C. elegans} from the corresponding frame is displayed in a color-coded depth image in Fig. \ref{fig_worm}c, showing the potential of our method to locate the neurons of the whole animal simultaneously in 3D. The full video is available in Visualization 1. The randomness of the diffuser also enables compressed sensing reconstructions with more voxels in the 3D result than pixels on the sensor. From a 4.2 mega pixel sensor, the reconstructed \textit{C. elegans} volume contains $50 \times$ more voxels and the gain could increase to $140 \times$ if we deconvolve with all the available PSFs within the $\SI{280}{\micro\meter}$ depth range. With regards to the resolvable voxels, for our experimental system the lower bound equals the imaging volume divided by the worst lateral ($\SI{2.9}{\micro\meter}$) and axial ($\SI{4}{\micro\meter}$) resolution, which gives 10 mega resolvable voxels per frame.

\section{Discussion and conclusion}
Like light field microscopy, our Fourier DiffuserScope achieves single-shot 3D imaging with high light throughput. Like Fourier light field microscope, it has efficient computation and reduced artifacts near the objective's native focal plane. Beyond Fourier light field microscope, our Fourier diffuser design and sparsity-based inverse algorithm enables nearly uniform resolution across a large imaging volume. Here, we: (1) provide a theoretical design framework for calculating system performance (e.g. resolution, volumetric FOV) from the diffuser parameters (e.g. number of microlenses, focal lengths), (2) carry out theoretical and numerical comparison to demonstrate that our RMM can achieve more uniform resolution across a larger imaging volume than a uni-focal MLA (Fourier light field microscope) or a RUM.

In the future, our system can be further improved in several ways. (1) For fabricating diffusers, our current indentation method is fast and cheap, but imprecise. While as-built surface shape should be captured by the PSF calibration and computationally accounted for, using more time-consuming and expensive manufacture methods (e.g. diamond turning, injection molding, or two-photon polymerization) could improve the surface quality of the diffuser to precisely fabricate a pre-defined diffuser surface and guarantee the system performance. (2) In our forward model, we didn't take into account scattering or use any space-time models for video processing. Particularly for neural activity tracking applications, the temporal behavior of calcium indicators can be used as a constraint~\cite{paninski2016temporal}, and the scattering potential can be incorporated to enable deeper imaging with higher-fidelity reconstructions, further suppressing noise and enhancing resolution. (3) Our first-principle derivation for our random diffuser design is made for a general-purpose imaging situation; however, for a specific type of data set, the microlenses locations and focal lengths distributions can be optimized using data-driven approaches for end-to-end learning. 


\section*{Supplemental materials}

\begin{figure}[h!]
\centering\includegraphics[width=\linewidth]{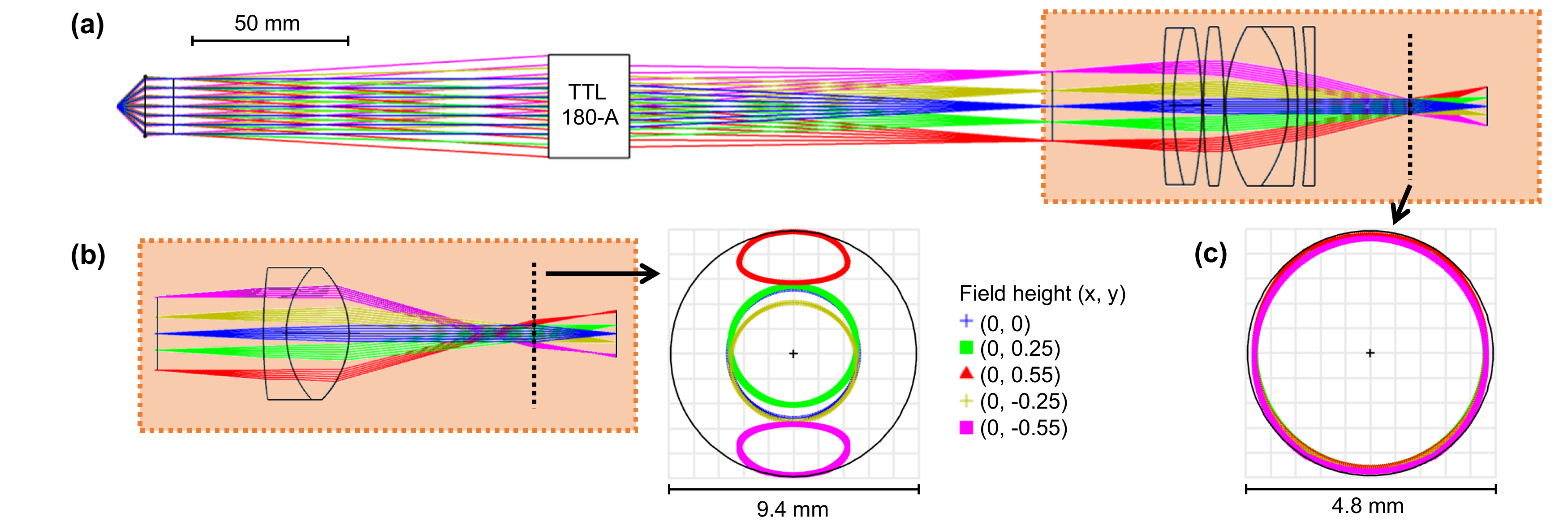}
\caption{Optimized design of the relay lens in Zemax OpticStudio. The optimized lens set consisting of off-the-shelf elements is shown in (a) orange box, and in (c) its resulting footprints from different field points mostly overlap. However, if a single achromatic doublet with similar focal length is used as the relay lens shown in (b) orange box, the footprints in (b) contains huge aberration and the shift-invariance assumption does not hold.}
\label{fig_zemax}
\end{figure}

In this section, we examine the design of the relay lens. The relayed pupil plane cannot have huge aberrations, otherwise the shift-invariant assumption does not hold. To investigate how aberration affects out system, we model the system in Zemax OpticStudio (Fig. \ref{fig_zemax}). The objective is represented by a paraxial plane since its lens data is not publicly available. The tube lens data is a black box model downloaded from Thorlabs (Thorlabs TTL180-A). There are two restrictions in choosing the relay lens: the focal lengths should be $< 60$ mm to make the full FOV recorded and the clear aperture is at least $\sim 30$ mm to prevent vignetting. However, we see a huge amount of aberration with the off-the-shelf lenses, even achromatic doublets.  To demonstrate, Fig. \ref{fig_zemax} (b) shows the layout of an achromatic lens with $50$ mm focal length (Edmund Optics 89682, clear aperture $39$ mm) and the resulting footprint of the relayed pupil plane at different field heights. Since the phase mask is put at the relayed pupil plane, the drifting footprints mean that different areas on the diffuser will be utilized for different field locations, which breaks the shift-invariant forward model. To implement an aberration-free and cost-effective relay lens, we designed a customized lens set with off-the-shelf elements. Noticing that the function of the relay lens is similar to an eyepiece in a traditional microscope, we start from the Erfle eyepiece design due to its wide FOV and short working distance \cite{eyepiece}. Since the Erfle contains one piece of uncommon convex-concave doublet, we replace it with a convex-convex achromatic doublet plus a concave-plano lens. We alternately optimize the radii of all the surfaces and replace every lens with the most similar counterpart in the catalog. We also optimize the air gap between every two components which can be controlled by using a spacers inside the lens tube. After many iterations, we arrive at the final design described in the Table \ref{table_lensdata}. In Fig. \ref{fig_zemax} (c), the new design's footprint of the peripheral field mostly overlaps with the center footprint and the aberration is greatly reduced. The relayed pupil size is $4.8$ mm which means that the effective de-magnification rate of the back pupil plane is $3.75 \times$ and the effective focal length of the lens set is $48$ mm. 

\begin{table}[h]
\begin{center}
\begin{tabular}{|p{3.5cm} |p{3cm}|p{2cm}|} 
\hline
\textbf{Lens model}& \textbf{Lens type} & \textbf{Focal length} \\ 
\hline
Edmund Optics 49-286 & Achromatic Doublet & 200mm \\ 
\hline
\multicolumn{3}{|c|}{Air gap 0.5 mm } \\
\hline
Newport KBX163 & Bi-Convex Lens & 175 mm \\ 
\hline
\multicolumn{3}{|c|}{Air gap 0.5 mm} \\
\hline
Thorlabs AC508-075-A & Achromatic Doublet & 75 mm \\ 
\hline
\multicolumn{3}{|c|}{Air gap 3 mm} \\
\hline
Newport KPC064 & Plano-Concave Lens &  $-$500 mm\\ 
\hline
\end{tabular}
\caption{\label{table_lensdata} Lens prescription table.}
\end{center}
\end{table}

\section*{Funding}
Gordon and Betty Moore Foundation (GBMF4562); Office of Naval Research (N00014-17-1-2401); David and Lucile Packard Foundation.

\section*{Acknowledgments}
The authors would like to thank Dr. Saul Kato and Vaishnavi Madhavan for providing the microfluidic chips and \textit{C. elegans} samples.

\section*{Disclosures}
The authors declare no conflicts of interest.

\bibliography{sample}

\end{document}